\documentstyle[epsfig,rotate]{ioplppt}

\begin{document}
\jl{4}

\title{ Physics of Strange Matter}

\author{Carsten Greiner\dag\ftnote{3}{Invited talk at
the IV International Conference on
'Strangeness in Quark Matter', Padova (Italy), July 20 - 24, 1998}}

\address{\dag\ Institut f\"ur Theoretische Physik, Universit\"at Giessen,
D-35392 Giessen, Germany}

\begin{abstract}
Relativistic heavy ion collisions offer the
possibility to produce exotic metastable states of nuclear matter
containing (roughly) equal number of strangeness compared to
the content in baryon number.
The reasoning of both their stability
and existence, the possible distillation of
strangeness necessary for their formation and the chances
for their detection are reviewed. In the later respect emphasize is put
on the properties of small lumps of strange quark matter with
respect to their stability against strong or weak hadronic decays.
In addition, implications in astrophysics like
the properties of neutron stars and the issue of baryonic dark matter
will be discussed.
\end{abstract}

\section{Introduction}

All known normal nuclei are made of the two nucleons, the proton and the neutron.
Besides those two lightest baryons there exist still a couple of other
stable (but weakly decaying) baryons, the hyperons.
Up to now the inclusion of multiple units of strangeness in nuclei remains
experimentally as theoretically rather largely unexplored.
This lack of investigation reflects
the experimental task in producing nuclei containing (weakly decaying)
strange baryons, which is conventionally limited by replacing one
neutron (or at maximum two) by a strange $\Lambda$-particle in scattering
experiments with pions or kaons.
There exists nowadays a broad knowledge about single hypernuclei,
i.e. nuclei, where one nucleon is substituted
by a $\Lambda $ (or $\Sigma $) by means of the exchange reaction
$\pi ^{+} + n \rightarrow \Lambda + K^+$.
Over the last two decades a rich phenomology has resulted for such hypernuclei.

However, there exist more or less no experimental insight how more than one
hyperon behave inside a nuclei. The technical
problem is to create within a tiny moment, smaller than the decay time
of a hyperon, enough hyperons and then to bring them together with nucleons
to form any potential multihypernucleus.
By employing a relativistic shell model calculation,
which gives a rather excellent description of normal
nuclei and single $\Lambda $-hypernuclei, it was
found that such configurations might exist as (small) bound
multihypernuclei (MEMO - metastable exotic multihypernuclear object)
\cite{Greiner,Sch92}.

Strange matter could also be realized in a completely different picture.
Indeed, this second and much more speculative possibility was raised
by physicists much earlier. The fundamental theory of strong interactions,
quantum chromodynamics, does not forbid the principle existence of `larger'
hadronic particles, so called multiquark states. Today only the
mesons and baryons are known in nature. However, there could exist states with
more than three quarks. Going further with this speculation one comes
to the conclusion that only multiquark states with nearly the same number
of up, down and strange quarks might exist as (meta-)stable configurations
\cite{Greiner,Fa84}. Such a very speculative form
of strange matter is called strange quark matter.

(Ultra-)relativistic heavy ion collisions provide
the only (earth based) source for the formation of either
strangelets (small lumps of strange quark matter) or multi-hypernuclear
objects, consisting of nucleons,
$\Lambda $'s and $\Xi $'s, as dozens of hyperons are produced in a single
central event. In principle, strangelets can be produced via two different
scenarios: by a condensation out of a quark-gluon plasma or by a coalescence
of hyperons out of the created hot and dense fireball. For the former scenario
it is essential that within the phase transition of the deconfined matter
to hadronic particles the
{\em net} strangeness (counting a surplus of strange over antistrange
quarks) is getting enriched in the plasma phase. This distillation
(or separation) of strangeness, i.e.
the possible conglomeration of net strangeness,
has been predicted to occur for a first order phase transition of a
baryonrich QGP \cite{CG1,Greiner}.
In particular, if the strangelet does exist in principle, it has to be regarded
as a cold, stable and bound manifestation of that phase being a remnant
or `ash' of the originally hot QGP-state. On the other hand
a further necessary 'request' for the possible condensation is
that the initially hot plasma phase has to cool down considerably
during the ongoing phase transition. Within our present knowledge
of the phase transition such a behaviour can neither be unambiqously
shown to happen nor be excluded.

In section 2 we briefly summarize the reasons for the (possible)
existence of this novel and exotic states.
In section 3 the mechanism of
strangeness distillation and the possible production
of small strange matter states are reviewed.
We conclude this section by discussing the detection
possibilities of small and finite strangelets with
respect to their lifetimes against strong or weak hadronic decays.
In section 4 we finally sketch on how the physics
of strange matter can affect the physical picture of dense neutron stars
and the issue of baryonic dark matter.

\section{Strange Matter}

\subsection{Strange quark matter}

The first speculation about the possible existence of collapsed
nuclei was given by Bodmer \cite{Bo71}. He argued that another form of
baryonic matter might be more stable than ordinary nuclei. Indeed it was
speculated there both on the possible existence of hyperonic matter with
baryons as colorless constituents or strange quark matter with quarks
as major constituents. The paper, however, lacked detailed calculation
as the MIT bag model or Walecka model were only available a few years later.

Let us now briefly summarize how a stable or metastable
strangelet might look like \cite{Greiner,Fa84}:
Think of bulk objects, containing a large number of
quarks $(u...u\, , \, d...d\, , \, s...s)$, so-called multiquark droplets.
Multiquark states consisting only of u- and d-quarks must have a mass
larger than ordinary nuclei, otherwise normal nuclei would be unstable.
However, the situation is different for
droplets of SQM, which would contain approximately the same
amount of u-, d- and s-quarks.
Speculations on the stability of strangelets are based on the following
observations:
(1) The (weak) decay of a s-quark into a
d-quark could be suppressed or forbidden because
the lowest single particle states are occupied.
(2) The strange quark mass can be lower than the Fermi energy of the u-
or d-quark in such a dense quark droplet. Opening a new flavour degree of
freedom therefore tends to lower the Fermi energy and hence also the mass per baryon of the
strangelet. SQM may then appear as a nearly neutral state.

\begin{figure}[ht]
\vspace*{\fill}
\centerline{\psfig{figure=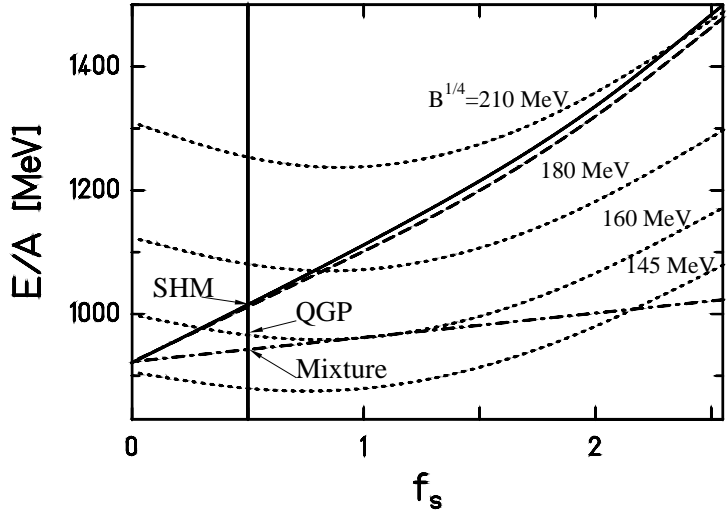,height=3.5in}}
\caption{
For a given strangeness fraction $f_s$ (number of strange quarks per
baryon number) the specific energy $E/A$
is shown for non-interacting cold quark matter for
different bag parameters and $m_s=150$~MeV and those of infinite
hyperonic matter. A tangent construction reveals that strong nucleon
decay  will move a strangelet candidate towards the tangent point.
\label{SQM}}
\end{figure}

If the mass of a strangelet is smaller than the mass of the
corresponding ordinary nucleus with the same baryon number, the strangelet
would be absolutely stable and
thus be the true groundstate of nuclear matter \cite{Fa84}.
Within the MIT bag model very low phenomenological
bag parameters $B^{1/4} \leq 150$ MeV have to be employed for modeling
such a possibility.
On the other hand, it is also conceivable that the mass per baryon
of a strange droplet is
lower than the mass of the strange $\Lambda $- baryon, but larger than the nucleon
mass (see Fig. 1). The droplet is then in a metastable state, it cannot decay
spontanously into
$\Lambda $'s.
For bag parameters
B$^{1/4}$ lower than 190 MeV strange quark droplets can only decay
via weak interactions.
For larger B-values strangelets are instable.
Due to the strong finite size effects
\cite{Greiner,Fa84,CG1} and the wider range of the employed model parameters,
smaller strangelets are much more likely to be metastable than
being absolutely stable. Also the possible influence of color
magnetic and color electric potentials has been necglected in these
calculations due to their complicated group structure. For very light
strangelets with baryon number $A<6$ it turns out \cite{Fa84}
that such contributions are in fact repulsive,
making the possible candidates less stable (with the well-known exception of
the H dibaryon).

\subsection{Multihypernuclear objects -- MEMOs}

\begin{figure}[ht]
\vspace*{\fill}
\centerline{\psfig{figure=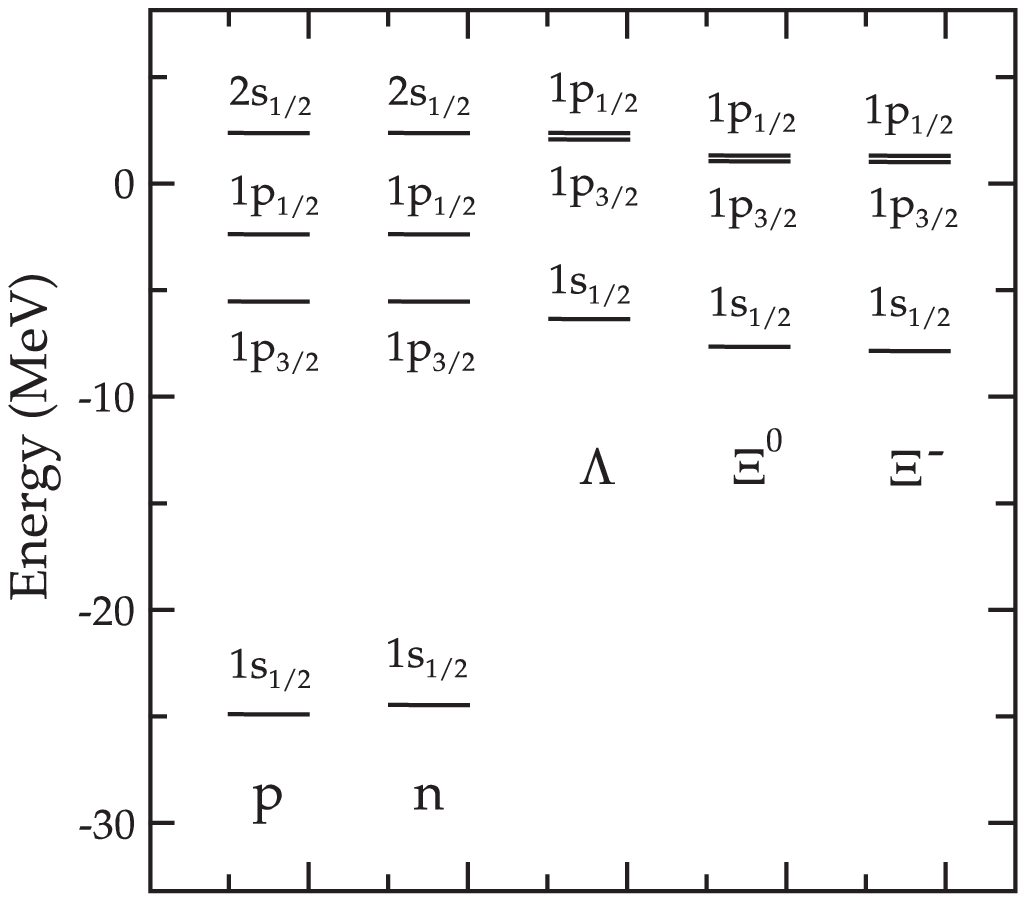,height=2.5in}}
\caption{
Single particle energy of a MEMO consisting of two of
each baryon of the baryon octet except the $\Sigma$'s.
The binding energy difference cancels the mass difference
of the strong reaction channels so that the whole 'nucleus' is metastable.
\label{pauli}}
\end{figure}

A classification scheme for metastable combinations of nucleons and
hyperons exhibits that combinations of nucleons,
$\Lambda$'s and $\Xi$'s are favoured compared to combinations with
$\Sigma$'s due to their Q-values in vacuum \cite{Greiner,Sch92}.
An example is shown in Fig. 2,
where the single particle levels of a strange nucleus consisting of
two of each proton, neutron, $\Lambda$, $\Xi^0$, $\Xi^-$ are plotted.
Note that each baryon sits in the 1s-state. The reaction
$\Xi N \to \Lambda \Lambda$
can not induce a strong decay because the two $\Lambda$'s sitting in
the 1s-level cause the produced $\Lambda$'s to escape in vacuum.
But this is energetically unfavoured resulting in an overall
metastable compound system.
The calculation was carried out in a relativistic mean field
model taking care of the nucleon-nucleon and nucleon-hyperon
interaction.

An extension of this model also implements
the scarce information about the hyperon-hyperon interaction.
This is done by introducing two new meson fields into the theory, 
$\sigma^*$ and $\phi$,
which couple to strange baryons only \cite{Greiner,Sch93}.
The binding energy for MEMOs will be moderately enhanced
(on a scale of nuclear binding energies) due to this additional
interactions. Indeed, binding energies of $-21$~MeV and more have
been found with a net strangeness fraction of $f_s\approx 1$.
Even negatively charged strange nuclear systems are possible without
loosing stability. The lightest stable
object of this type is likely to be $(2\Lambda+2\Xi^0+2\Xi^-)$.

The global properties of a
strangelet or a MEMO are likely to be identical:  a
similar small charge $|Z|$ and nearly the same average baryon density $\sim 3 \rho _0$.
In turn, this would suggest that a nearly neutral and heavy candiate
could not unambiguosly be considered as a strangelet.
In principle, to distinguish experimentally between both one has in
addition to resolve the mass $E/A$ very accurately. A MEMO is only bound in the
order of $E_B/A\sim 10 $ MeV whereas the strangelet may be bound from
$10 - 200$ MeV (which is, of course, speculation).
The resembling microscopic structure gives raise to the speculation
that both states might have a strong overlap and correlation.
The MEMO would decay into a strangelet, if the latter is energetically more
favourable.

\section{Production possibilities in relativistic heavy ion collisions}

Multiple collisions in the reaction of two bombarding nuclei
ensure that the interacting system starts to equilibrate which might be suited to search
for the most interesting collective effects.
In particular all exotic objects need for their
formation
{\em large strange particle numbers,
 high degree of
equilibration} and
{\em large densities.}
In order to get this kind of states one should therefore use
high energies to produce enough strangeness and energy density,
and heavy nuclei to
gain as much equilibration as possible. This situation
is now achieved at Brookhaven AGS using Au+Au and at CERN SPS using Pb+Pb.
It may then be possible to produce some of the lightest
multistrange objects in the laboratory.

\subsection{Strangeness distillation and strangelet condensation}

In the following we want to sketch why the production of
SQM clusters, if they do exist in principle, is
likely, if a baryon rich and hot deconfined QGP is created in such collisions.
The net strangeness of the QGP is zero from the onset,
although an equal, however large, number of strange and antistrange
quarks has been produced.
However, there is a physical mechanism which separates the strange quarks
from their antiparticles \cite{CG1} (see Fig. 3a).
It is `simple' for the antistrange quarks to materialize
in kaons K($q\bar s$) because of the lots of light quarks
as compared to the s-quarks which could only move into the suppressed antikaons
($K^-$) or the heavy hyperons.
Hence, during a near equilibrium phase
transition a large antistrangeness builds up in the hadron matter
while the QGP retains a large net strangeness excess.

\begin{figure}[ht]
\vspace*{\fill}
\centerline{\psfig{figure=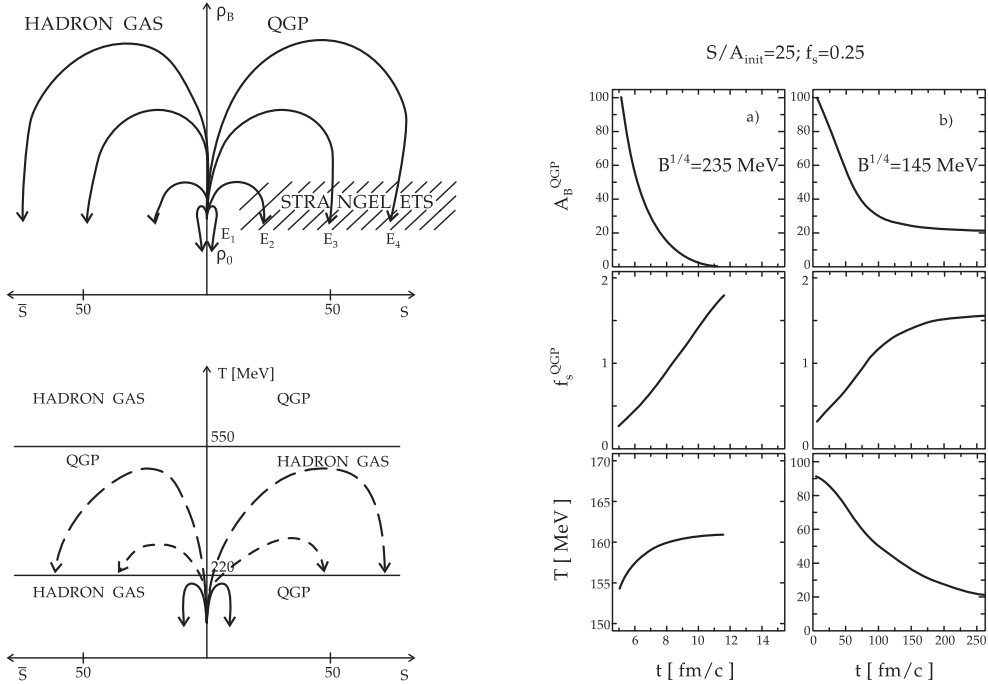,height=4.0in}}
\caption{
(Left) The fountain of strangeness production. The distillation of strangeness
is more effective for high baryon densities.
(Right) a) Baryon number, strangeness content and temperature of the
quark glob during complete hadronization as a function of time
for a very large bag constant
$B^{1/4}=235$ MeV.
Note the strong increase of the strangeness content with time.
b) The same situation as in a), however, for a small bag constant
$B^{1/4}=145 $ MeV, when a strangelet condenses.
One observes a strong decrease in the evolving temperature.
\label{QGP}}
\end{figure}

For modeling the evolution of an initially hot fireball a two phase
equilibrium description between the hadron gas and the QGP was combined
with nonequilibrium radiation by incorporating the rapid freeze-out
of hadrons from the hadron phase surrounding the QGP droplet during
a first order phase transition (last reference of \cite{CG1}).

Two scenarios may describe the evolution to the final state:
The quark droplet may remain unstable until the
strange quarks have clustered 
into $\Lambda $-particles and other strange hadrons to carry away 
the strangeness 
and the plasma has completely vanished into standard particles.
This scenario is customarily accepted.
However, if SQM exists at low temperatures
in configurations having a mass per baryon 
lower
than the mass of the $\Lambda $-particle, the hot SQM
droplet would remain at the phase transition
boundary much longer.
As shown in \cite{CG1},
producing strange baryons like
$\Lambda$ particles
is energetically more
expensive and therefore less likely than producing SQM
like strangelets. Towards
the end of the evolution only baryons
are allowed to escape from the droplet, since at this point all of the 
antiquarks are gone. The baryons will be mostly nucleons,
since the hyperons are heavier and require more energy for
formation. These nucleons remove energy but they do not 
carry away any strange
quarks, so the ratio of strange to nonstrange quarks increases
further, refining the distillation of strangeness.
The hot strange matter in fact might
cool down to cold lumps of size $A_B \sim 5-50$, depending on the
original baryon content of the plasma.

Fig. 3b gives an impression
how the hadronisation proceeds for a large bag constant
($B^{1/4}=235$ MeV -- no strangelet in the groundstate)
and a small bag constant ($B^{1/4}=145$ MeV).
For the large bag constant the system hadronizes completely in
$\Delta t \sim $ 8 $\frac{fm}{c}$, which is customarily expected and thus
not too surprising.
Yet, a strong increase of the net strangeness of the system
is found in both situations,
and
the plasma drop reaches a strangeness fraction of
$f_s \sim  1.5 $ when the volume becomes small.
Indeed, for the small bag constant, however,
a {\em cold} strangelet emerges from the expansion and evaporation process
with an approximate baryon number of $A_B \sim$  17, a radius of
$R \sim  2.5  fm$, and a net strangeness fraction of
$f_s(t \rightarrow  \infty ) \stackrel{>}{\sim }  1.5$, i.e. a charge to baryon
ratio $Z/A = (1-f_s)/2
\sim  - 0.2$!

If a baryon rich QGP is temporarily created it will assemble
the strange quarks during the ongoing evolution and expansion.
It might either lead to the formation of a strangelet, or it will decay mainly into hyperons
in its late stage. The strangelet would be a remnant or `ash' of the deconfined
quark matter phase and thus would provide a signal for the formation of
a QGP \cite{CG1}.

As emphasized already in the introduction, for the possible condensation
of small pieces of strange quark matter out of an initially hot QGP
with no net strangeness two mechanisms have to be realized by nature:
(1) strangeness distillation and (2) ongoing cooling (e.g.
by pion evaporation) during the phase transition resulting in a loss
of internal heat.

The strangeness distillation during the hadronization of a baryonrich QGP
is a very intuitive process. It works for all model parameters like
e.g. also much higher bag constants \cite{Sp98}. In a recent
study the above described model including particle evaporation
from the outer surface has been combined within 1-dimensional
longitudinal hydrodynamical expansion \cite{Du97} with initial conditions
expected at CERN-SPS energies and future RHIC energies. Again, a strong
accumulation of net strangeness in the plasma phase
has been found during the ongoing phase transition.
Still, the mechanism by which a QGP state is converted into hadrons is
a major uncertainty in the different descriptions.
The hadronisation transition has often been
described by  geometric and statistical models,
where the matter is assumed to be in partial or complete equilibrium
during the whole expansion phase. A fully microscopic
and numerical model of the hadronization
of confined and color-neutral hadrons
out of a deconfined plasma has very recently been realized
within the Friedberg-Lee model by Traxler et al. \cite{Tr98}.
Exploratory studies of a baryonrich system within this model
have shown in fact that the strangeness distillation is a genuine feature
of the hadronisation. It might, however, be very difficult to experimentally
verify that this distillation indeed happens during the transition.
We remark that the separation mechanism might be probed
by density interferometry with hyperons or kaons
\cite{CG89}. The hadrons with negative strangeness, the
$\Lambda $, $\bar{K}$ and $\bar{K}^0$, are expected to be produced
mainly at the last stage of the phase transition when the size of
the quark phase volume has become quite small.

The realization of the second mechanism by nature,
i.e. the necessary ongoing cooling for final
possible strangelet condensation, is not as obvious.
It was pointed out
already in the second ref. of \cite{CG1}
that the strangelet formation can only go hand in hand with
strong cooling rather than reheating.
Besides the expansion of the system additional
pion and nucleon evaporation should help to
allow for a possible, yet necessary cooling.
On the other hand one realizes from Fig. 3b that depending on
the bag parameter employed there might be
(moderate to strong) cooling in the one and reheating in the other case.
In all the numerous and recent studies within thermal hadronic models
one tries to find common parameters
of the fireball by trying to 'fit' the ratios of measured hadronic abundancies
within a few parameters. Within the above model of a rapidly disintegrating
QGP we had repeated these kind of analysis \cite{Sp98} and were forced
to assume a high bag constant of $B^{1/4} \approx 210 - 235 $ MeV
in order to find a similar good agreement like earlier investigations.
Then, according to the model, no real cooling would emerge during the
phase transition and thus also no strangelet may condense.
The temperature will drop, if
the specific entropy per baryon in the hadron phase
exceeds that in the quark phase,
$(S/A)^{HG} > (S/A)^{QGP}$,
otherwise the temperature has to slightly increase during the transition
\cite{CG1}.
The first is the case (within the model)
when the bag constant $B^{\frac{1}{4}}$ is small or moderate and allows for
the existence of (meta-)stable strange quark matter states.
Although this intimate connection between cooling of the
plasma phase and the existence of strange quark matter is intriguing,
it might be valid only within the simple parametrisation of the
quark gluon plasma phase within a bag model description.
Ultimately, whether
$(S/A)^{HG}$ is larger or smaller than $(S/A)^{QGP}$ at finite, nonvanishing
chemical potentials might theoretically
only be proven rigorously by lattice gauge calculations
in the future, as also the principle existence of (meta-)stable strange quark
matter.

\subsection{Production via coalescence}

More conservative estimates of production of small multistrange objects
(either strangelets or MEMOs) without the need of a temporarily present
intermediate QGP phase are based on coalescence models being put
forward by Dover and coworkers \cite{Do94}.
Such estimates yield very small
production rates, for instance $3\times 10^{-9}$ events per central
Au+Au collisions at 11.7~AGeV for the lightest bound $\Xi$ system
$_{\Xi^0\Lambda\Lambda}^7$He.
Very simple coalescence estimates give production probabilities
of strange clusters of the order of $10^{3-A_B-|S|}$, where $S$ denotes the
strangeness and $A_B$ the baryon number of the cluster.
For a maximum sensitivity of $\approx 10^{-10}$
only strangelets or MEMOs with baryon numbers
of $A_B \le 8$ are expected to be seen.
As has been seen recently by the E864-collaboration \cite{Sa98}
the penalty factor $q$ for an additional unit of baryon number
at AGS energies in central collisions is in fact very small, $q\approx 1/50 $,
implying that the formation of exotic objects by coalescence is
even less favorable.

\begin{figure}[ht]
\vspace*{\fill}
\centerline{\psfig{figure=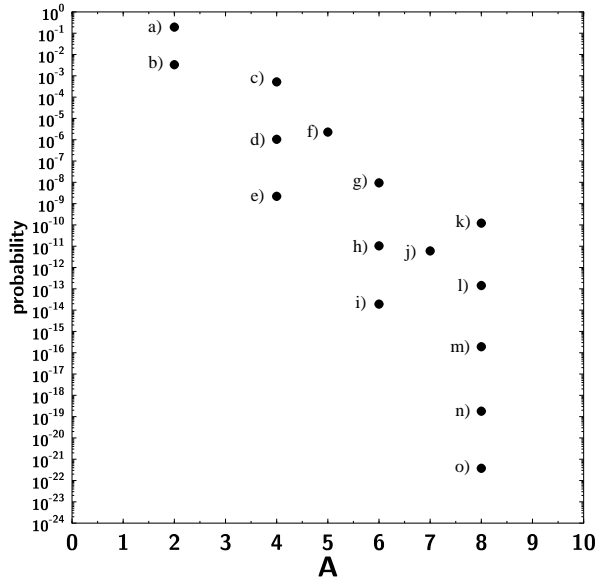,height=3.5in}}
\caption{Calculated multiplicities of hypermatter clusters from a 
hadronizing QGP with
$A_B^{init}=416$, $S/A^{init } =40$, $f_s^{init }=0$ and a bag constant of
$B^{1/4} =235$~MeV:
a) $H^0$ ($m_{H} =2020$~MeV),
b) \{ $\Xi^-,\Xi^0$ \} ,
c) ${}^4He$,
d) \{ $4\Lambda$ \} ,
e) \{ $2\Xi^-,2\Xi^0$ \} ,
f) $^5_\Lambda He$,
g) $^{\;\;6}_{\Lambda\Lambda} He$,
h) \{ $2n,2\Lambda,2\Xi^-$ \} ,
i) \{ $2\Lambda,2\Xi^0,2\Xi^-$ \} ,
j) $^{\;\;\;\;7}_{\Xi^0\Lambda\Lambda} He$,
k) $A=8$, $S=0$, 
l) $A=8$, $S=-4$,
m) $A=8$, $S=-8$, 
n) $A=8$, $S=-12$, 
o) $A=8$, $S=-16$
($-S$ gives the number of strange quarks).
}
\end{figure}

Fig. 4 shows calculated thermal multiplicities of various hypermatter
clusters for central Pb+Pb collisions at SPS energies \cite{Sp98}.
These numbers should serve as an upper limit for the production via
coalescence. The (thermal) penalty factor
suppresses the abundances of heavy clusters:
metastable hypermatter can only be produced with a probability $p< 10^{-8}$
for $A \ge 4$ (e.g. a \{$2\Xi^-,2\Xi^0$\} object). 

Hence, only exotic objects with very low mass number
are expected to be (possibly) seen at the AGS or at CERN.

\subsection{Decay channels of small strangelets and detection possibilities}

It is important to note that these objects
are a new form of matter,
not a specific new particle. The strange droplets produced in
these reactions do not
come in the form of a single type of particle.
Many different sizes of droplets may be produced, spanning
a range in mass, charge, and strangeness content. 
The experimental task of finding the new form of matter 
is therefore challenging.
Here any detected particle having
an unusual charge to mass ratio is a potential strange matter
candidate.

\begin{figure}[th]
\centerline{\psfig{figure=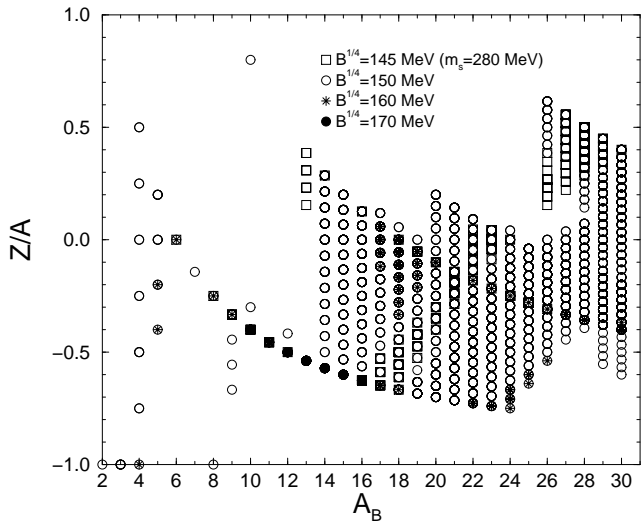,height=8cm}}
\caption{The charge fraction $Z/A$ 
for long-lived strangelets, which are stable against 
strong and weak hadronic decay, for different choices of the bag parameter.
The case for the original MIT bag model 
parameters ($B^{1/4}=145$ MeV, $m_s=280$
MeV) is also plotted.}
\label{fig:stab30}
\end{figure}

Employing TOF-techniques in present settings
to reveal the velocity and thus the charge to mass ratio,
the experimental setup sets a natural time scale of $\approx 50 $ ns
\cite{Sa91,Cr91,Pr93}.
So, an important question we finally have to adress are the lifetimes of these
objects.
The lifetime of a MEMO should be similar to the $\Lambda $'s lifetime,
i.e. $\sim 10^{-10}$ sec.
However, all the present
experiments are unable to observe metastable hyperclusters
due to the required lifetimes $\tau \gg 10^{-10}$s.
On the other hand,
if a produced strangelet is absolutely stable,
the only
energetically possible decay mode is the rather slow weak leptonic decay ($s \rightarrow
d$, $Q \rightarrow  Q'+ e + \bar{\nu }$),
which will turn the
strangelet to its minimum value in energy.

It is much more likely, if at all, however, that a small strangelet
is a weakly decaying metastable state.
The situation turns out to be
even more complicated:
Strangelets will not be in their ground state
when being produced in a heavy ion collision. Suppose a strangelet
is created out of the hot and dense matter with some arbitrary strangeness,
charge and baryon number. Strong interactions and the distillation
process will alter the composition of a strangelet
by particle evaporation on a timescale of a few hundred fm/c after
the collision. The strangelet, if surviving this, will cool down
until it reaches the domain of weak interactions. Weak hadronic decay
by hadron emission takes place on a timescale being estimated between
$10^{-7} - 10^{-9}$ s. As a conservative quideline one might even also consider
the life-time of the $\Lambda $-particle.
Strangelets stable against further strong decay channels but decaying by
the weak hadronic decay will be dubbed as short-lived candidates.
Strangelets
stable against weak hadronic decay will then still be subject to the
weak leptonic decay occuring on a timescale of $10^{-4} - 10^{-5}$ s.
They are dubbed as long-lived candidates.
Obviously, the present experiments are probably only capable to look
for the long-lived candidates.

Following the qualitative and old ideas of Chin and Kerman \cite{Ch79},
we had carried out a detailed analysis for small
short-lived and long-lived strangelet candidates \cite{Sch97,CG1}, where
shell effects are quite pronounced and immportant.
One qualitative outcome is that
all candiates are likely to be negative, especially the
long-lived candiadates.  At first sight this might appear
counter-intuitive. A closer inspection reveals that strong nucleon decay
will drive a strangelet to a higher net strangeness content $f_s$ and
carries away positive charge.
Consider a system of bulk SQM
in the ground-state with a finite total strangeness
of e.g. $f_s=0.4$ (compare Fig. 1, $B^{1/4}=160 $ MeV).
One would now naively argue that strange quark matter with $f_{s}$=0.4
is the ground-state of the system,
because the energy per baryon of this state is lower than
that of the hyperonic
state. However, the total energy per baryon can be lowered by
additional $\sim$50 MeV by
assembling the non-strange quarks into pure
nucleonic degrees of freedom (i.e. `$\alpha $-particle or nucleon emission'),
leaving the strange quarks in a
strange matter droplet, its strangeness fraction enriched to $f_{s} \approx 1$
slightly {\em above} its minimum value.
This strong nucleon decay
will stop to happen for energetical reasons
at the so called tangent point (compare Fig.1).
Finite size corrections, i.e. strong shell effects,
on the other side, of course,
will change this picture quantitatively. A similar reasoning shows
that the weak nucleon decay again will change the candidates to
become even more negative.

Fig. 5 shows the result of this investigation for long-lived
candidates which are stable against weak hadronic decay modes \cite{Sch97}.
Different bag parameters have been chosen in order to get some feeling
for the different possibilities.
Besides the {\em neutral} so called quark-alpha state \cite{Mi88} with
$N_u=N_d=N_s=6$ ($A=6$) a valley of stability appears at $A=8 \, - \, 16$
for long-lived, negative candidates.
For bag parameters of $B^{1/4} \geq 180$ MeV
no long-lived candidates have been found at all.

Still there exist a much
richer spectrum of shortlived strangelets or MEMOs with a lifetime
of the order of the $\Lambda $ or somewhat below \cite{Sch97,CG1}.
It might well be that only metastable strange
clusters with $\sim$ cm flight path seem to have a chance of being created.
Future experiments geared to
proof the (non)existence of strangelets therefore should clearly cover such
short lifetimes.
An open geometry detectional device will
be needed, which
clearly is a challenging task due to the large background of charged hadrons
at the target in violent events with high multiplicities.

\section{Strange Matter in Astrophysics}

\subsection{'Neutron' stars}

Originally (strange) quark matter in bulk was thought to exist
only in the interiour of neutron stars where the pressure is
high enough that the neutron matter melts into its quark substructure \cite{SMProc}.
At least in the cores of neutron stars (where the density rises up to the order of $10$ 
times normal nuclear density)
it is not very likely that matter consists of individual hadrons. 

On the other hand it is also known that the pure `neutron' matter
is not really a nuclear matter state made solely out of neutrons, but
at least at higher densities consists also of a considerable amount of
protons as well as hyperons.
Indeed, it was shown by Glendenning that
hyperons \cite{Gl85} appear at a moderate density of about
$2 \, - \, 3$ times normal nuclear matter density.
These new species influence the properties of the equation of state
of matter and the global properties of neutron stars.
There may be so many hyperons in the neutron star that the whole object is more
appropriately dubbed a giant hypernucleus.

The gross structure of a neutron star like its mass M and radius R is
influenced by the composition of its stellar material. 
This holds especially in the case of the existence of strangeness bearing 
``exotic'' components like hyperons or strange quark matter
which may significantly change the
characteristic mass-radius (MR) relation of the star.
For example hyperons considerably soften the EOS and reduce the maximum mass
of a neutron star.

\begin{figure}[ht]
\vspace{0.5cm}
\[\mbox{\tt figure6.gif}\]
\vspace{0.5cm}
\caption{Pressure surfaces: A = pure hadronic phase at low densities;
MP = mixed phase of both hadronic and quark matter phase, B =
pure quark phase at very high baryon densities
(bag constant: $B^{1/4}=165\,MeV$, compression modulus of hadronic EOS: $K=300 MeV$)}
\label{phasetransition}
\end{figure}

Besides the speculative possibility of nearly pure SQM stars, which would be the case,
if SQM is absolutely stable in bulk,
it is more likely that in the interiour of the star
a phase transition to hadronic matter will take place.
The `neutron' star would then have the form of a
so called hybrid star,
i.e. a star which is made of baryonic matter in the outer region, but with
a quark matter core in the deep interiour.
The deconfinement phase transition from hadronic matter to the SQM phase 
is constructed according to the requirement
of global charge neutrality between both phases \cite{KS97}.
The pressure of both
phases spans up a two dimensional surface over the plane of the
relevant chemical potentials
of the baryons ($\mu_n$) and the electrons ($\mu_e$).
This situation is depicted in Fig. 6.
One finds that the
pressure varies smoothly and continuously with the proportion
of both phases in equilibrium, which, in turn
will lead to a mixed phase of finite radial extent
(of several kilometers) inside the star \cite{KS97}.

\begin{figure}[ht]
\centerline{\epsfig{file=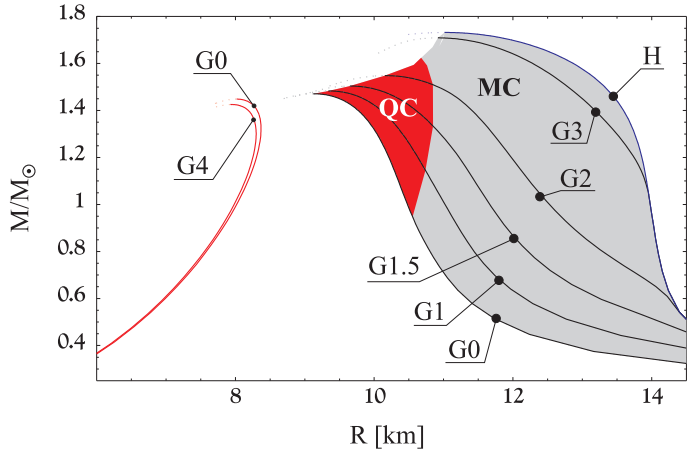,height=6cm}}
\caption{Mass radius relation for pure SQM stars ($R<9km$) and hybrid stars 
($R>9km$), H= pure hadron, QC=star has a pure quark core, MC=star has a mixed
phase core
($B^{1/4}=165\,MeV$,
$K=300 MeV$)}
\label{RM1}
\end{figure}

Fig. 7 shows the resulting MR relations for various situations and QCD coupling
constants $g$ employed within the effective mass bag model
discussed in \cite{KS97}.
The left hand side ($R<9\,km$) shows the pure SQM star results, which
turn out to be rather compact objects.
The situation
changes completely for the hybrid stars, depicted on the right hand side of
fig.\,\ref{RM1} ($R>9\,km$). Their
MR relation approaches the curve of the pure hadron
star (denoted by H) if the phase transition occurs deep insight the star.
Accordingly it will be difficult to judge from measured properties of pulsars
to really disentangle the possible interiour structure of 'neutron stars'.
For possible consequences which might be observable, like e.g.
timing structure of pulsar spin-down, higher spin rates, or stronger
cooling rates, especially relevant for young 'neutron' stars,
we want to refer to \cite{Lu98}.

\subsection{Dark matter}

Witten raised the intriguing possibility that
strange quark matter might in principle also
be absolutely stable and may also provide an
explanation for cold (baryonic) dark matter in the universe \cite{Fa84}.
If being stable and nearly neutral, it could exist
at all possible sizes \cite{Ruj84}, as the small Coulomb energy
is not sufficient for a break up into smaller pieces \cite{Fa84}.
It might thus span as a charge neutral state the empty `nuclear desert' \cite{Ruj84}
within the range
from $A_B \sim 300$ up to sizes of neutron stars $A_B \sim 10^{56}$.

In Witten's prospective scenario hot strange quark matter
nuggets could have condensed out of the
deconfined phase during the phase transition in the expanding and
cooling early universe \cite{Fa84}.
These would carry most of the tiny surplus in baryon number of the whole universe.
The baryon number would remain inside if the heat and entropy of the nugetts is
carried away mainly by neutrino emission instead of mesons and especially
baryons. If an absolutely stable groundstate would exist
the hot nuggets would further cool and instead of suffering a complete hadronisation
they might settle into these new states and hence could resolve the
dark matter problem. Since then the idea of absolute stability has
stimulated a lot of work on potential consequences in astrophysics
\cite{SMProc}. Today it is fair to say that there is still no real evidence
for absolutely stable SQM to be a possible dark matter candiate.
There had been early conjectures \cite{SMProc} that escpecially smaller
glumps of SQM could not have survived the ongoing and still hot epoch of
the cosmological evolution, but would have in fact been
completely evaporated into single hadrons. Hence, particle physicists were
led to look for other possible (and maybe even more exotic) explanations
on the issue of dark matter. On the other side, in a
very recent (and speculative) study \cite{En98} the issue of stable quark matter
was taken up again. There it is speculated that a (stable) pre quark-matter
phase might in fact also explain the formation of galaxies clusters,
the formation of massive black holes in the galactic centres, and also
might be a source of the $\gamma $-ray bursts. We refer the interested reader
to \cite{En98}.
\vspace*{1em}
\\
{\bf  Acknowledgements:}

Thanks go to  C. Dover, A. Dumitru, A. Gal, L. Gerland,
P. Koch-Steinheimer, D. Rischke, J. Schaffner-Bielich, K. Schertler,
C. Spieles, H. St\"ocker, M. Thoma and C. Traxler for
their help and collaboration.
%%

%
%----------------------------------------------------------------------%
%                             End of Text                              %
%----------------------------------------------------------------------%

\section*{References}

\begin{thebibliography}{99}

\bibitem{Greiner} for a general review see:
    C. Greiner and J. Schaffner,
    {\it Int. J. Mod. Phys.}{\bf E5} (1996) 239;
    C. Greiner and J. Schaffner, 'Physics of Strange Matter', nucl-th/9801062 ;
    \\
    for an introduction see:
    H. Crawford and C. Greiner, `The Search for Strange Matter',
    {\it Scientific American}, Vol. 270 No. 1, January 1994, p. 58
\bibitem{Sch92}
    J. Schaffner, C. Greiner and H. St\"ocker, {\it Phys. Rev. C }{\bf 46}, 322 (1992)
\bibitem{Fa84}
    E. Witten, {\it Phys. Rev. D }{\bf 30}, 272 (1984);
    E. Farhi and R. L. Jaffe, {\it Phys. Rev. D} {\bf 30}, 2379 (1984)
\bibitem{CG1}
    C. Greiner, P. Koch and H. St\"ocker, {\it Phys. Rev. Lett. }{\bf 58}, 1825 (1987);
    C. Greiner, D.H. Rischke, H. St\"ocker and P. Koch, {\it Phys. Rev. D }{\bf 38}, 2797 (1988);
    C. Greiner and H. St\"ocker, {\it Phys. Rev. D }{\bf 44}, 3517 (1991)
\bibitem{Bo71}
    A. R. Bodmer, {\it Phys. Rev. D }{\bf 4}, 1601 (1971)
\bibitem{Sch93}
    J. Schaffner, C. B. Dover, A. Gal, C. Greiner and H. St\"ocker,
    {\it Phys. Rev. Lett.} {\bf 71}, 1328 (1993); J. Schaffner,
    C. Dover, A. Gal, C. Greiner, D. Millener and H. St\"ocker,
    {\it Ann. Phys. }{\bf 235}, 35 (1994)
\bibitem{Sp98}
    C. Spieles, C. Greiner and H. St\"ocker, {\it Eur. Phys. J.} {\bf C 2}, 351 (1998)
\bibitem{Du97}
    A. Dumitru, C. Spieles, H. St\"ocker and C. Greiner,
    {\it Phys. Rev. }{\bf C 56}, 2202 (1997)
\bibitem{Tr98}
    C. Traxler, T.S. Biro and U. Mosel,
    'Hadronization of a Quark-Gluon Plasma in the Chromodielectric Model',
    hep-ph/9808298 ;
    C. Traxler and C. Greiner, in preparation
\bibitem{CG89}
    C. Greiner and B. M\"uller, {\it Phys. Lett. }{\bf B219}, 199 (1989);
    M. Gyulassy, {\it Phys. Lett. }{\bf B286}, 211 (1992);
    A. Jacholkowski, contribution to these proceedings
\bibitem{Do94}
    A.J. Baltz, C.B. Dover, S.H. Kahana, Y. Pang, T.J. Schlagel, E. Schnedermann,
    {\it Phys. Lett. }{\bf B325}, 7 (1994)
\bibitem{Sa98}
    J. Sandweiss, contribution to these proceedings;
    Z. Xu, contribution to these proceedings;
    G. van Buren, contribution to these proceedings;
    M. Munoz, contribution to these proceedings
\bibitem{Sa91}
    J. Sandweiss, {\it Nucl. Phys. B }(Proc. Suppl.) {\bf 24B},
    234 (1991); \\
    F.S. Rotondo, Proc. Quark Matter `96,
    {\it Nucl. Phys. A }{\bf 610}, 297c (1996)
\bibitem{Cr91}
    H.J. Crawford et al., {\it Nucl. Phys. B }(Proc. Suppl.) {\bf 24B},
    251 (1991)
\bibitem{Pr93}
    F. Dittus, Proc. Quark Matter `95,
    {\it Nucl. Phys. A }{\bf 590}, 347c (1995);
    R. Klingenberg, Proc. Quark Matter `96,
    {\it Nucl. Phys. A }{\bf 610}, 306c (1996)
\bibitem{Ch79}
    S. A. Chin and A. K. Kerman,
    {\it Phys. Rev. Lett. }{\bf 43}, 1292 (1979)
\bibitem{Sch97}
    J. Schaffner, C. Greiner, A. Diener and H. St\"ocker,
    {\it Phys. Rev. }{\bf C 55}, 3038 (1997);
    J. Schaffner, 'Strangelets and Strange Quark Matter', nucl-th/971104
\bibitem{Mi88}
    F. C. Michel, {\it Phys. Rev. Lett. }{\bf 60}, 677 (1988)
\bibitem{SMProc}
     for a detailed survey on the astrophysical implications see:
     `Strange Quark Matter in Physics and Astrophysics',
     edited by J. Madsen and P. Haensel,
     {\it Nucl. Phys. B }(Proc. Suppl.) {\bf 24B}
\bibitem{Gl85}
    N.K. Glendenning, {\it Astrophys. J.}{\bf 293}, 151 (1985);
    F. Weber and N.K. Glendenning, {\it Astrophys. J. }{\bf 390}, 541 (1992);
    J. Schaffner and I. Mishustin, {\it Phys. Rev. C }{\bf 53}, 1416 (1996)
\bibitem{KS97}
    K. Schertler, C. Greiner and M.H. Thoma, {\it Nucl. Phys. A }{\bf 616}, 659 (1997);
    K. Schertler, C. Greiner, P. Sahu and M.H. Thoma,
    'The Influence of Medium Effects in the Gross Structure of Hybrid Stars',
    astro-ph/9712165, to appear in {\it Nucl. Phys. A}
\bibitem{Lu98}
    N. Glendenning, S. Pei and F. Weber, {\it Phys. Rev. Lett. }{\bf 79}, 1603 (1997);
    N. Glendenning, {\it J. Phys. G}: Nucl. Part. Phys. {\bf 23}, 2013 (1997);
    T. Lu, 'Observational Effects of Strange Stars', astro-ph/9807052;
    J. Madsen, 'How to identify a strange star', astro-ph/9806032
\bibitem{Ruj84}
     A. De Rujula and S.L. Glashow, {\it Nature }{\bf 312}, 734 (1984);
     A. De Rujula, {\it Nucl. Phys. A}{\bf 434}, 605c (1985)
\bibitem{En98}
    D. Enstr\"om, S. Fredriksson, J. Hanson, A. Nicolaidis and S. Ekelin,
    'A Quark-Matter Dominated Universe', astro-ph/9802236
%
%
\end{thebibliography}
\end{document}